\documentclass[12pt]{article}
\usepackage{geometry}
\usepackage[utf8]{inputenc}
\usepackage{natbib}
\usepackage{authblk}
\usepackage{url}
\usepackage{color}
\usepackage{graphicx}
\usepackage{longtable}
\usepackage{setspace}
\usepackage{changepage}
\usepackage[normalem]{ulem}
\usepackage{arydshln}

\usepackage{chngcntr}

\newcommand{\red}[1]{\textcolor{black}{#1}} 
\newcommand{\blue}[1]{\textcolor{black}{#1}}

\usepackage{amsmath,amssymb,amsthm} 

\title{Simulating near-field enhancement in transmission of airborne viruses with a quadrature-based model}
\author[1]{Laura Fierce}
\author[1,2]{Alison Robey}
\author[1,2]{Cathrine Hamilton}
\affil[1]{Environmental \& Climate Sciences Department, Brookhaven National Laboratory}
\affil[2]{Science Undergraduate Laboratory Internship Program, Department of Energy Office of Science}
\date{}

\begin{document}

\maketitle

\section*{Abstract}
Airborne viruses, such as influenza, tuberculosis, and SARS-CoV-2, are transmitted through virus-laden particles expelled when an infectious person sneezes, coughs, talks, or breathes. These virus-laden particles are more highly concentrated in the expiratory jet of an infectious person than in a well-mixed room, but this near-field enhancement in virion exposure has not been well quantified. Transmission of airborne viruses depends on factors that are inherently variable and, in many cases, poorly constrained, and quantifying this uncertainty requires large ensembles of model simulations that span the variability in input parameters. However, models that are well-suited to simulate the near-field evolution of respiratory particles are also computationally expensive, which limits the exploration of parametric uncertainty. In order to perform many simulations that span the wide variability in factors governing transmission, we developed the Quadrature-based model of Respiratory Aerosol and Droplets (QuaRAD). QuaRAD is an efficient framework for simulating the evolution of virus-laden particles after they are expelled from an infectious person, their deposition to the nasal cavity of a susceptible person, and the subsequent risk of initial infection. We simulated 10,000 scenarios to quantify the risk of initial infection by a particular virus, SARS-CoV-2. The predicted risk of infection was highly variable among scenarios and, in each scenario, was strongly enhanced near the infectious individual. In more than 50\% of scenarios, the physical distancing needed to avoid near-field enhancements in airborne transmission was beyond the recommended safe distance of two meters (six feet)\blue{ if the infectious person is not wearing a mask}, though this distance defining the near-field extent was also highly variable among scenarios. \blue{We find this variability in the near-field extent is explained predominantly by variability in expiration velocity.} These findings suggest that, during outbreaks of airborne viruses, it is best to maintain at least three meters of distance to avoid local increases in virion exposure near an infectious person; protections against airborne transmission, such as N95 respirators, should be available for work conditions where distancing is not possible.

\section{Introduction}
Airborne transmission of pathogens was first recognized over fifty years ago \citep{riley1959} and has since been acknowledged as an important transmission route for a number of infectious diseases \citep{yu2004,bloch1985,zhao2019,kim2016,leclair1980}. Even so, the risks posed by airborne spread remain poorly addressed by many safety standards \citep{bahl2020}, a trend that continued with dire consequences during the early phases of the COVID-19 pandemic \citep{morawskaandmilton2020}. It is now apparent that transmission of SARS-CoV-2, the virus causing COVID-19, occurs primarily through the airborne route \citep{samet2021,prather2020}. Though long-range airborne transmission is possible due to the small size and long transport distances of many infectious respiratory particles, the risk of infection is greatest near infectious individuals where the concentration of virions is highest \citep{meyerowitz2020}. These near-field enhancements in transmission must be understood to design and implement effective mitigation strategies, such as selecting the distance needed between desks in a classroom or between workers in a factory. However, near-field increases in virion exposure are subject to variable and poorly constrained parameters, and this uncertainty in factors controlling transmission is not easily represented in models.

A key challenge in quantifying near-field effects is the inherent variability in factors governing transmission risk. For example, the rate at which different individuals expel virions varies by orders of magnitude \citep{chen2020, leung2020}, and epidemiological data suggests that the risk of infection at a given virion dose is also highly variable \citep{chen2020vari}. These uncertainties in the physiological characteristics of both infectious and susceptible individuals leads to large uncertainty in predictions of transmission risk \citep{gale2020, watanabe2010}. Large ensembles of model simulations are needed to quantify this uncertainty. However, while computational fluid dynamics models simulate air flows within a room in great detail and are, therefore, are able to resolve heterogeneity in virion concentrations \citep[e.g.][]{beghein2005, choi2012}, these models are computationally expensive \citep[e.g.][]{beghein2005, choi2012}, limiting the number of simulations that can be performed. Multi-zone models that simulate air flow within buildings typically track the average concentration in each room \citep[e.g.][]{li2005, emmerich1994}; by assuming virions are well mixed within each room, these models neglect local increases in virion concentrations near an infectious person. To represent airborne transmission across a wide range of conditions, a model framework is needed that accurately represents processes governing near-field and far-field transmission but is efficient enough to perform large ensembles of simulations. 

To address this need, we introduce the Quadrature-based model for Respiratory Aerosol and Droplets (QuaRAD), an efficient framework for simulating the full life cycle of virus-laden particles within indoor spaces --- from their initial creation and expulsion from an infectious person to their eventual removal from the space or their deposition in the nasal passages of a new host. This model framework is an alternative to Monte Carlo models for simulating the dispersion and evolution of respiratory particles. In contrast to Monte Carlo models, which represent particle size distributions using thousands of randomly sampled particles, quadrature-based moment methods have been shown to accurately represent low-order distribution moments --- as well as other integrated quantities --- using only a small number of quadrature points \citep{mcgraw1997,fierce2017}.  In this study, the quadrature representation was used to quantify the rate at which particles deposit into the most likely initial infection site of SARS-CoV-2, the nasal epithelium. This paper describes the QuaRAD model (Section \ref{sec:model_description}) and its application to simulation across a wide range of scenarios (Section~\ref{sec:results}). Though we focus here specifically on SARS-CoV-2, most aspects of this model are equally applicable and easily adaptable to studying the airborne transmission of any disease.

\section{Model description}\label{sec:model_description}
The respiratory particles that carry airborne viruses are created in the respiratory system through expiratory activities such as sneezing, coughing, talking, or breathing. A particle's size at the time it is expelled dictates the distance it travels before settling to the ground. If a particle is transported to another individual and inhaled, its size influences if and where in the respiratory system it is most likely to deposit. For this reason, the QuaRAD model is designed to accurately and efficiently represent the size distribution of expelled particles and its evolution through evaporation, transport, and removal. 

\begin{figure}
\centering
\includegraphics[width=6.in]{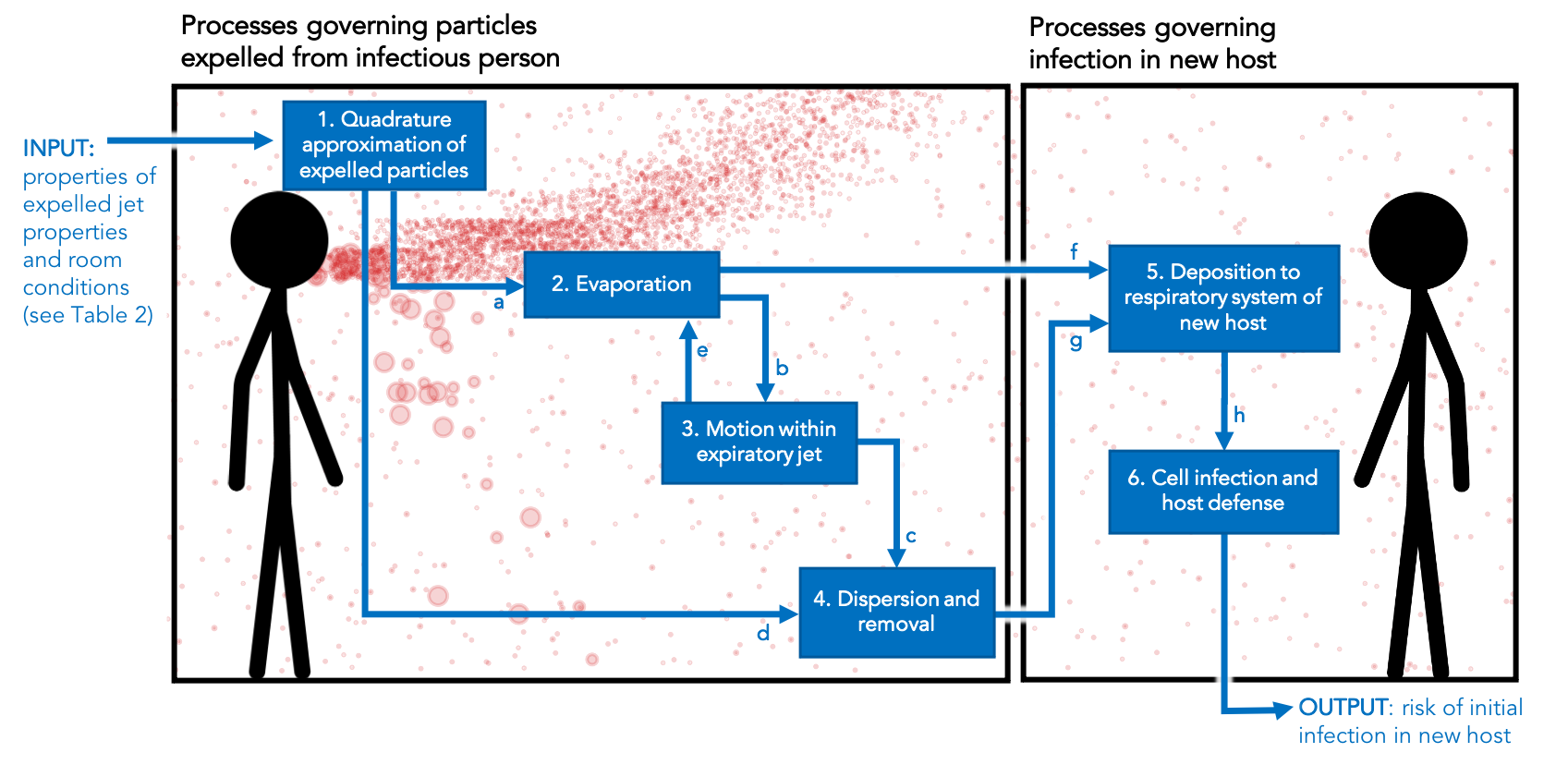}
\caption{QuaRAD represents processes governing the evolution of expelled particles (left) and processes governing infection in a new host (right). Model inputs are provided in Table~\ref{tab:inputs}. The model components (1--6) and the connections components (a--h) are described in the text of Section~\ref{sec:model_description}.}
\label{fig:schematic} 
\end{figure}

The processes simulated in QuaRAD are shown in Fig.~\ref{fig:schematic}. The model represents mechanisms governing the transport and evolution of virus-laden particles (boxes 1--4), as well those governing virion deposition and initial infection within the potential host (boxes 5--6). Before describing the model components in detail, we first provide an overview of the connections between model components (labeled with letters in Fig.~\ref{fig:schematic}).

Each of our QuaRAD simulations begins by replacing the continuous size distribution of expiratory particles with a quadrature approximation consisting of six weighted particles (see Section~\ref{sec:quadrature}). Taking as inputs (a) the initial diameter of each quadrature point, as well as the chemical properties of the constituent aerosol, we then simulate water evaporation to predict particle sizes as they shrink over time (see~\ref{sec:evaporation}). Using the (b) time-dependent size of each particle, we then predict how it would move within the expiratory jet if it were released at the jet center; the (c) time-dependent position of the center line particles and the (d) quadrature weights are taken as inputs to a model that represents particle dispersion using a Gaussian puff within a turbulent jet (see Section~\ref{sec:dispersion}). The (e) velocity of the particle relative to the gas is fed back to the evaporation scheme to represent enhancements in evaporation rates through convection. 

The output of the particle dispersion model is used to predict deposition \blue{to the nasal cavity of a new host and the probability of} initial infection. Using the (f) size of each quadrature point and the (g) virion concentration associated with each quadrature point, we predict the total virion dose reaching the nasal epithelium of a susceptible person as a function of their location relative to the infectious person and the duration of the encounter. From the (h) virion dose, we then predict the probability that each deposited virion finds an ACE2 receptor, binds to it, and begins replication  (Section~\ref{sec:infection}). For convenience, all variable used in this section are also defined in Table~\ref{tab:def}.

\subsection{Quadrature approximation of particle size distributions}\label{sec:quadrature}
QuaRAD uses numerical quadrature to obtain an accurate and efficient representation of the size distribution of expelled aerosol particles and droplets. To construct the quadrature approximation, we follow the approach of \citet{johnson2011} and represent expired particles through three different mechanisms, resulting in three lognormal distributions. Breath and speech create near continuous flows of small aerosol particles through the bronchial fluid film burst (BFFB) mechanism, the process of thin mucus films rupturing as the alveoli open and close deep in the lung (referred to as bronchial or b-mode particles); activities that vibrate the vocal cords further create slightly larger particles through laryngeal vibrations (referred to as laryngeal or l-mode particles); and expulsion of mucus from the mouth or nose generates the largest particles (referred to as oral or o-mode particles). 

In QuaRAD, each of the three lognormal distributions is approximated using Gauss-Hermite quadrature. We found that one-point, three-point, and two-point quadrature were sufficient for representing the b-, l-, and o-modes, respectively, such that the overall distribution is represented with only six weighted particles; adding additional particles did not improve simulation accuracy (see Appendix~\ref{ap:quad_optimization}). 

Gauss-Hermite quadrature approximates integrals of the form $\int_{-\infty}^{\infty}\exp(-x^2)f(x)dx$; with a change of variables, integrals over lognormal distributions can also be expressed in this form. The expected value of a function of the natural logarithm of the diameter, $f(\ln D_{\text{p}})$, over a continuous lognormal distribution is given by:
\begin{equation} \label{eqn:expval}
E[f(\ln D_{\text{p}})] = \int_{-\infty}^{\infty}\frac{1}{\sigma\sqrt{2\pi}}\exp\bigg(-\frac{(\ln D_{\text{p}}-\mu)^2}{2\sigma^2}\bigg)f(\ln D_{\text{p}})d\ln D_{\text{p}},
\end{equation}
where $\mu$ is the natural logarithm of the geometric mean and $\sigma$ is the natural logarithm of the geometric standard deviation, the parameters of the lognormal distribution. With the following change of variables:
\begin{equation}
x = \frac{\ln D_{\text{p}} - \mu}{\sqrt{2}\sigma}, 
\end{equation}
the integral over the continuous distribution is estimated by a quadrature approximation of the form:
\begin{equation}
E[f(\ln D_{\text{p}})] \approx \frac{1}{\sqrt{\pi}}\sum_{i=1}^nw_if(\sqrt{2}\sigma h_i + \mu),
\end{equation}
where the abscissas $h_i$ are the roots of the physicists' Hermite polynomial: 
\begin{equation}
H_n(x)=(-1)^n\exp\big(x^2\big)\frac{d^n}{dx^n}\exp\big(-x^2\big), 
\end{equation}
and the weights are given by:
\begin{equation}
w_i=\frac{2^{n-1}n!\sqrt{\pi}}{n^2[H_{n-1}(h_i)]^2}
\end{equation}

The overall number concentration of virions, $N_{\text{v}}$, is then given as the sum over $N_{\text{v},i}$, \blue{the number concentration of} virions associated with each quadrature point: 
\begin{equation}
N_{\text{v}}=\sum_i^{N_{\text{quad}}}N_{\text{v},i}.
\end{equation}
The number of virions associated with a given quadrature point $i$ is computed as a function of the particle's viral load --- that is, the number of virions per volume of respiratory fluid --- and its initial diameter:

\begin{equation}
N_{\text{v},i}=N_{\text{p}}\frac{\pi}{6}D_{0,i}^3v_iw_i,
\end{equation}
where $N_{\text{p}}$ is the overall \blue{particle} number concentration and $v_i$ is the viral load associated with quadrature point $i$. 

\begin{figure}
\centering
\includegraphics[width=2.5in]{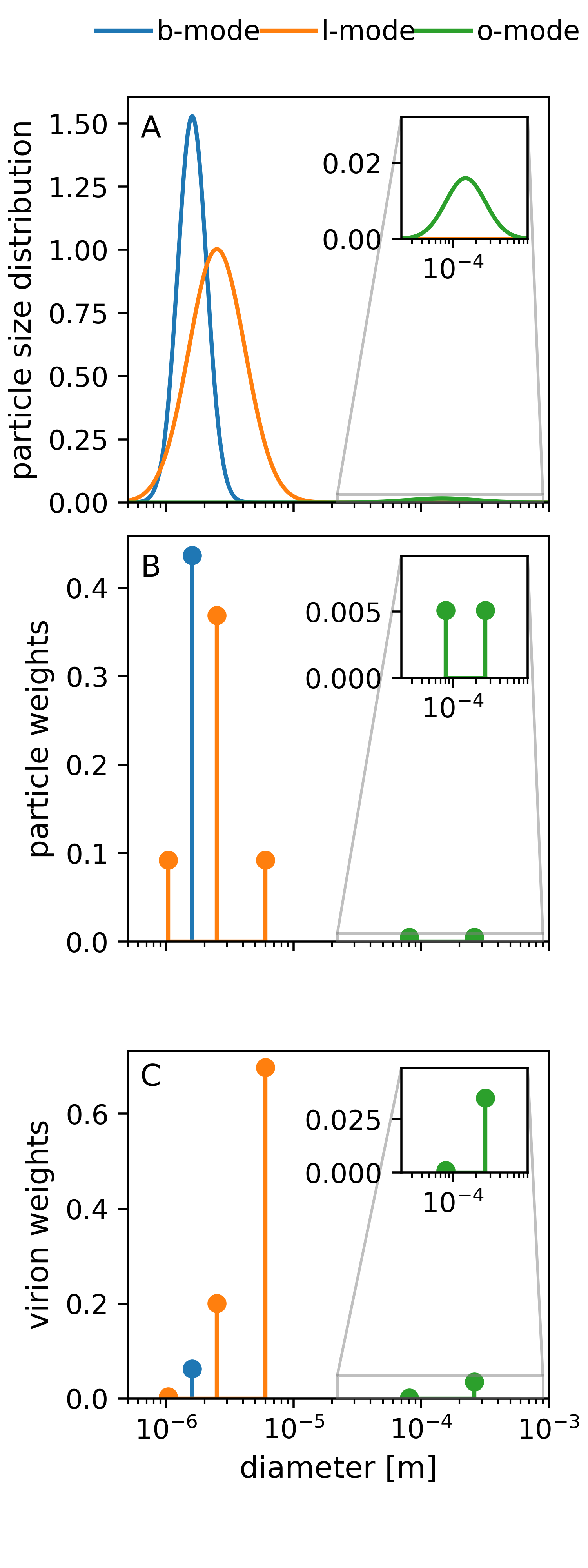}
\caption{Example of the particle size distribution represented as three separate lognormal modes of particles originating from the bronchial (blue), laryngeal (orange), or oral (green) region of the respiratory system. In QuaRA, the (A) continuous distributions for the b-, l-, and o-mode particles are represented using (B) 1-point, 3-point, and 2-point quadrature, respectively. Weights representing distributions of virions are shown in (C).}
\label{fig:quad_emission} 
\end{figure}

The distribution in virions with respect to particle size is an important parameter when modeling transmission, but the variation of SARS-CoV-2 loading within particles of different sizes has not \blue{yet} been well quantified. A small number of studies have investigated virion expiration rates for other viruses with comparable viral loads \citep{jacot2020} in fine and coarse particles using the Gesundheit II (G-II), an instrument designed to capture and analyze respiratory particles \citep{milton2013, leung2020}. We combine the measurements of viral shedding from the G-II with the measurements of the particle size distributions by \citet{johnson2011}, \citet{morawska2009}, and \citet{asadi2019} to estimate the viral load for particles in each mode. We assume that all virions in fine particles ($D_{\text{p}}<5$~$\mu$m) are distributed among the b- and l-mode particles and that virions in coarse particles ($D_{\text{p}}>5$~$\mu$m) are contained in o-mode particles; we assume the viral load is uniform across fine particles and across coarse particles. The weights corresponding to the distribution in virions, given by $N_{\text{v},i}/N_{\text{v}}$, are shown in Fig.~\ref{fig:quad_emission}c.

\subsection{Water evaporation from particles}\label{sec:evaporation}
Expelled aerosol particles and droplets are predominantly water and contain small amounts of dissolved or suspended aerosol components, such as mucin, salts, and surfactants \citep{vejerano2018}. After they are expelled, water evaporates, causing the particles to shrink \citep{wei2015, morawska2009}, which may impact the distance a particle is able to travel before settling through sedimentation \citep{xie2007} and its deposition efficiency to nasal epithelium \citep{cheng2003, heyder2004}.

When first expelled, we assume particles are suspended in air that is saturated with water vapor ($\text{RH}=100\%$) and at a temperature near the average human body temperature ($T_{0}=310.15$) \citep{wei2015}. Following measurements by \citet{vejerano2018}, we assume that freshly expelled particles and droplets are 1\%-9\% dry aerosol by volume and that the aerosol mixture has an effective hygroscopicity parameter $\kappa$ between 0.05 and 1. Once expelled, the particles cool and evaporate, with the rate of evaporation and equilibrium diameter dependent on the particles' initial size, water content, and the hygroscopic properties of its aerosol constituents.

Following the approach of \citet{kukkonen1989} and \citet{wei2015}, we modeled evaporation by solving the following coupled set of ordinary differential equations:
\begin{align}\label{eqn:evaporation}
&\frac{dm_{\text{p}}}{dt}=\frac{2\pi pD_{\text{p}}M_{\text{w}}D_{\infty}C_{\text{T}}\text{Sh}}{RT_{\text{v},\infty}}\ln\bigg(\frac{p-p_{\text{v,p}}}{p-p_{\text{v},\infty}}\bigg)\\
&\frac{dT_{\text{p}}}{dt}=\frac{1}{m_{\text{p}}C_{\text{p}}}\bigg(\pi D_{\text{p}}^2k_g\frac{T_{\text{v},\infty} - T_{\text{p}}}{0.5D_{\text{p}}}\text{Nu}-L_{\text{v}}\frac{dm_{\text{p}}}{dt}\bigg),
\end{align}
where $m_{\text{p}}$ is the mass of the aqueous particle, $T_{\text{p}}$ is the particle temperature, $p$ is the ambient pressure, $p_{\text{v,p}}$ is the vapor pressure at the droplet surface, $p_{\text{v},\infty}$ is the vapor pressure far from the droplet surface, $M_{\text{w}}$ is the molecular weight of water, $D_{\infty}$ is the binary diffusion coefficient of water vapor in air far from the droplet surface, $R$ is the universal gas constant, $C_{\text{p}}$ is the specific heat of the particle, $k_{\text{g}}$ is the thermal conductivity of air, $L_{\text{v}}$ is the latent heat of vaporization, $C_{\text{T}}$ is a correction factor, Sh is the Sherwood number, and  Nu is the Nusselt number. We assume spherical droplets to compute the diameter $D_{\text{p}}$ as a function of the mass of aerosol --- assumed constant over the simulation --- and water, which evolves according to Eqn.~\ref{eqn:evaporation}.

The evaporation model given in Eqn.~\ref{eqn:evaporation} includes the enhancement in evaporation rate due to convection. This effect depends on the Sherwood number, the Nusselt number, and a correction factor $C_{\text{T}}$ for the diffusion coefficient due to the temperature difference between the particle and the air:
\begin{equation}
C_{\text{T}}=\bigg(\frac{T_{\text{v},\infty}-T_{\text{p}}}{T_{\text{v},\infty}^{\lambda-1}}\bigg)\bigg(\frac{2-\lambda}{T_{\text{v},\infty}^{2-\lambda}-T_{\text{p}}^{2-\lambda}}\bigg),    
\end{equation}
where $\lambda$ is a constant between $1.6$ and $2$, fixed at $\lambda=1.6$ in this study. The Sherwood number is given by:
\begin{equation}
\text{Sh}=1+0.38\text{Re}^{1/2}\text{Sc}^{1/3},
\end{equation}
where Sc is the Schmidt number, given by:
\begin{equation}
\text{Sc}=\frac{\nu}{D_\infty},
\end{equation}
where $\nu$ is the dynamic viscosity. The Nusselt number is given by:
\begin{equation}
\text{Nu}=1+0.3\text{Re}^{1/2}\text{Pr}^{1/3},
\end{equation}
where Pr is the Prandtl number, given by:
\begin{equation}
\text{Pr}=\frac{C_{\text{p}}\mu}{k_{\text{g}}}.
\end{equation}

The rate at which an aqueous droplet evaporates is driven by the difference between the vapor pressure over the droplet and the vapor pressure of the ambient air, where droplet vapor pressures much larger than the ambient vapor pressure will result in rapid evaporation rates. We compute the vapor pressure over an aqueous particle using the Kappa-K\"{o}hler model \citep{petters2007}:
\begin{equation}
p_{\text{v,p}}=p_{\text{v},0}\bigg(\frac{D_{\text{p}}^3-D_{\text{d}}^3}{D_{\text{p}}^3-D_{\text{d}}^3(1-\kappa)}\bigg)\exp{\bigg(\frac{4\sigma_{\text{s/a}}M_{\text{w}}}{RT_{\text{p}}\rho_{\text{w}}D_{\text{p}}}\bigg)},
\end{equation}
where $p_{\text{v},0}$ is the saturation vapor pressure, $D_{\text{d}}$ is the particles' dry diameter, $\kappa$ is the effective hygroscopicity parameter of the aerosol contained in the particle, $\sigma_{\text{s/a}}$ is the surface tension at the particle surface, and $\rho_{\text{w}}$ is the density of water. The vapor pressure over an aqueous droplet is computed relative to the saturation vapor pressure over a flat surface of pure water, which is computed as a function of the vapor temperature $T_{\text{v}}$ according to the Buck equation \citep{buck1981}:

\begin{equation}
p_{\text{v},0}=611.21\exp{\Bigg(\bigg(19.84-\frac{T_\text{v}}{234.5}\bigg)\bigg(\frac{T_\text{v}-273.15}{T_\text{v}-16.01}\bigg)\Bigg)}.
\end{equation}

\begin{figure}
\centering
\includegraphics[width=3.in]{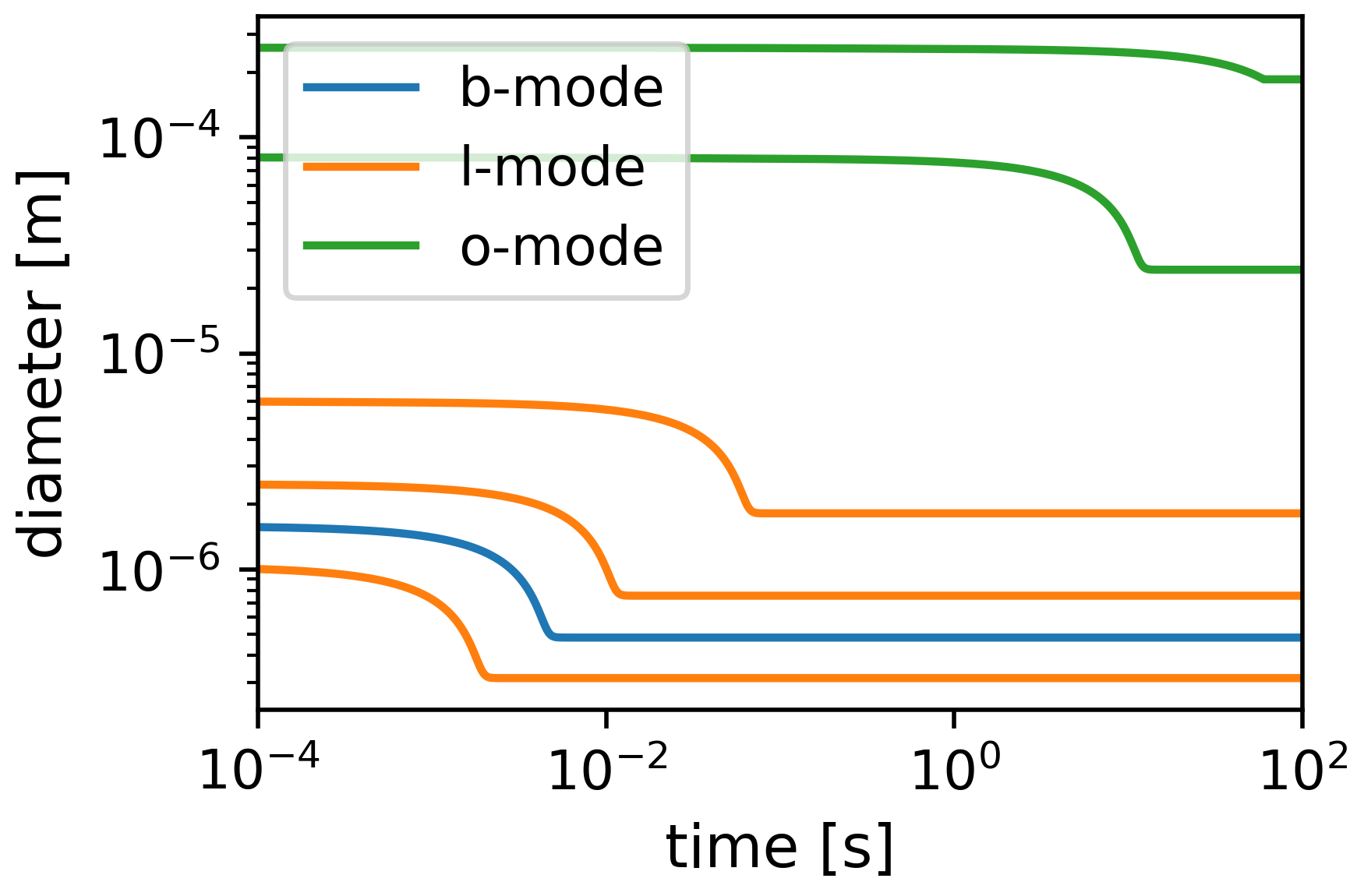}
\caption{Temporal evolution of the diameter of each quadrature point simulated in the baseline scenario. The diameters at $t=0$ correspond to their diameter at emission (shown in Fig.~\ref{fig:quad_emission}). Particles in the b- and l-mode, which tend to be on the order of tens of micrometers or smaller, reach their equilibrium size within $10^{-3}$ to $10^{-1}$~s after expulsion, depending on their size, whereas the large, o-mode particles evaporate more slowly and typically do not reach their equilibrium sizes before settling.}
\label{fig:Dp_vs_t} 
\end{figure}

For the example scenario, the evolution of $D_{\text{p},i}(t)$ is shown in Fig.~\ref{fig:Dp_vs_t} for each quadrature point $i=1,...,N_{\text{quad}}$\blue{, upon exposure to the room-averaged temperature and relative humidity}. All particles, regardless of their initial size, shrink over time, but the timescale for evaporation varies over orders of magnitude, as shown in previous studies \citep{morawska2006, redrow2011, wei2015}. The largest particles (o-mode, green) require more than ten seconds to reach their equilibrium size. On the other hand, the fine particles in the b-mode (blue) and l-mode (orange) reach equilibrium within $10^{-3}$ to $10^{-1}$ seconds, depending on their initial size. A particle's diameter influences its transport through the expiratory jet, as described in the following section.

\begin{figure}
\centering
\includegraphics[width=3.in]{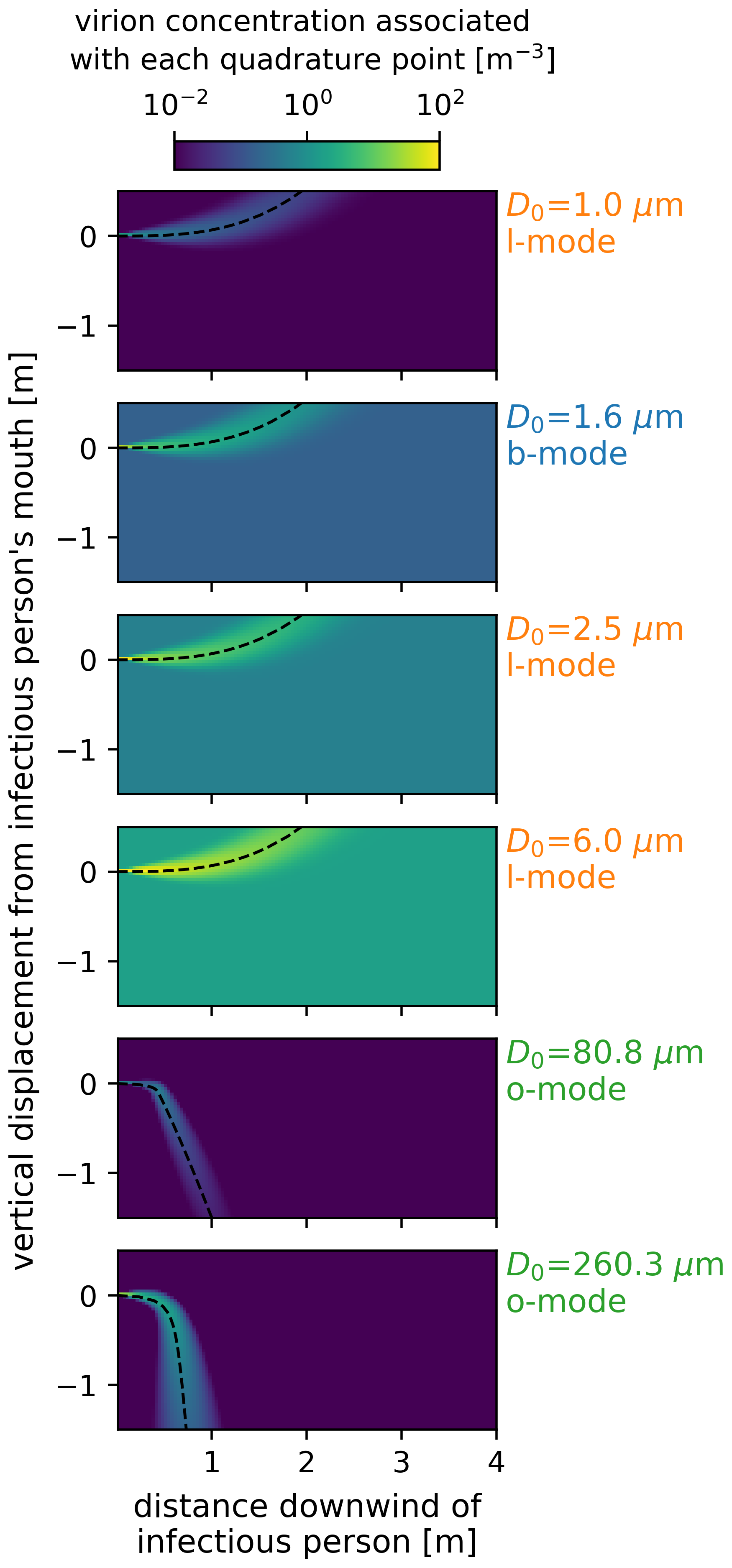}
\caption{The virion concentration $N_{\text{v},i}$ associated with each of the quadrature points $i=1,...,6$, shown in Fig.~\ref{fig:quad_emission}. Larger weights in Fig.~\ref{fig:quad_emission} correspond to higher virion concentration. Black dashed lines show the center line trajectory for each quadrature point, whereas the false color plots show the virion concentration, predicted using a Gaussian puff model of dispersion about the center line. The overall virion concentration is then computed as the sum over each quadrature point, shown for this baseline example in Fig.~\ref{fig:scenario__C_vs_xz}.}
\label{fig:scenario__Cq_vs_xz} 
\end{figure}

\subsection{Particle dispersion within a turbulent jet}\label{sec:dispersion}
To predict the number concentration of virus-laden particles at a position $(x,y,z)$, we represent the evolution of each quadrature point using the exposure model for indoor sources from \citet{drivas1996}. For an instantaneous source, such as a single cough, the concentration associated with each quadrature point $i$ is a function of position as well as time, denoted $c_i(x,y,z,t)$. A continuous emission source, such as talking, is represented as a series of puffs, such that the steady state concentration $\tilde{c}_i(x,y,z)=\int_{-\infty}^{\infty}c_i(x,y,z,t)dt$, as shown for each quadrature point in Fig.~\ref{fig:scenario__Cq_vs_xz}.

Following \citet{drivas1996}, we represent dispersion as the combination of a decay term and reflection terms $R_x(t)$, $R_y(t)$, and $R_z(t)$:
\begin{equation}
c_i(x,y,z,t) = \frac{q\exp{\bigg(-\bigg(\frac{\text{ACH}}{3600} - \frac{w_{\text{d}}A}{V}\bigg)t\bigg)}}{\pi^{3/2}b^3}R_x(t)R_y(t)R_z(t),
\end{equation}
where \text{ACH} is the ventilation rate in air changes per hour, $w_\text{d}$ is the deposition velocity, $A$ is the deposition area, and $V$ is the room volume. The reflection terms in \citet{drivas1996} account for reflections from the walls, floor, and ceiling, which are important for dispersion of gas-phase contaminants and small particles. We represent these wall reflections for particles smaller than $30~\mu$m, and $R_x(t)$, $R_y(t)$, and $R_z(t)$ are given by: 

\begin{align}
R_x(t)=\sum_{i=-\infty}^{\infty}\Bigg[\exp\bigg({-\frac{(x+2iL-x_{\text{c}}(t))^2}{b(t)^2}}\bigg) + \exp{\bigg(-\frac{(x+2iL+x_{\text{c}}(t))^2}{b(t)^2}\bigg)}\Bigg]\label{eqn:Rx}\\
R_y(t)=\sum_{i=-\infty}^{\infty}\Bigg[\exp{\bigg(-\frac{(y+2iW-y_{\text{c}}(t))^2}{b(t)^2}\bigg)} + \exp{\bigg(-\frac{(y+2iW+y_{\text{c}}(t))^2}{b^2}\bigg)}\Bigg]\label{eqn:Ry}\\
R_z(t)=\sum_{i=-\infty}^{\infty}\Bigg[\exp{\bigg(-\frac{(z+2iH-z_{\text{c}}(t))^2}{b(t)^2}\bigg)} + \exp{\bigg(-\frac{(z+2iH+z_{\text{c}}(t))^2}{b(t)^2}\bigg)}\Bigg]\label{eqn:Rz},
\end{align}
where $L$, $W$, and $H$ are the length, width, and height of the room, $(x_{\text{c}}(t),y_{\text{c}}(t),z_{\text{c}}(t))$ is the trajectory of a particle located in the center of the puff, and $b(t)$ is the puff width. We assume that the infectious person is standing in the center of the room ($x_0=L/2$,$y_0=W/2$). In Eqns.~\ref{eqn:Rx}--\ref{eqn:Rz}, each infinite sum is replaced as the sum from -1000 to 1000\blue{, which was sufficient for convergence}.

For particles larger than $30~\mu\text{m}$, we do not include reflections. The terms $R_x(t)$, $R_y(t)$, and $R_z(t)$ are reduced to:
\begin{align}
R_x(t)=\exp{\bigg({-\frac{(x-x_{\text{c}}(t))^2}{b(t)^2}}\bigg)}\\
R_y(t)=\exp{\bigg({-\frac{(y-y_{\text{c}}(t))^2}{b(t)^2}}\bigg)}\\
R_z(t)=\exp{\bigg({-\frac{(z-z_{\text{c}}(t))^2}{b(t)^2}}\bigg)}.\\
\end{align}

Whereas \citet{drivas1996} represents exposure from the release of a toxin under quiescent conditions, we represent particle dispersion during talking using a turbulent jet model that includes buoyancy and gravitational settling. In the original formulation, $x_{\text{c}}$, $y_{\text{c}}$ and $z_{\text{c}}$ are fixed at the location of the pollutant source. In contrast, we simulate the trajectory $x_{\text{c}}(t)$, $y_{\text{c}}(t)$, and $z_{\text{c}}(t)$ within the expiratory jet of the infectious person. We assume that breadth of the puff is the same as the breadth of the jet, which is proportional to the distance traveled along the center line of the jet, $s(t)$:
\begin{equation}
b(t)=0.114s(t).
\end{equation}

To compute the position and velocity of a particle with diameter $D_{\text{p}}$ at the center of each puff, we follow the general approach of \citet{wei2015}. The temporal evolution of particle position and velocity is determined by solving the set of six ordinary different equations that describe the motion of a particle in a moving gas \citep{crowe2011}:
\begin{align}
&\frac{dx_{\text{c}}}{dt}=u_{\text{p}}\\
&\frac{dy_{\text{c}}}{dt}=v_{\text{p}}\\
&\frac{dz_{\text{c}}}{dt}=w_{\text{p}}\\
&\frac{du_{\text{p}}}{dt}=\frac{3\rho_{\text{g}}C_{\text{D}}}{4D_{\text{p}}\rho_{\text{p}}}\big(u_{\text{g}} - u_{\text{p}}\big)|u_{\text{g}} - u_{\text{p}}|\\
&\frac{dv_{\text{p}}}{dt}=\frac{3\rho_{\text{g}}C_{\text{D}}}{4D_{\text{p}}\rho_{\text{p}}}\big(v_{\text{g}} - v_{\text{p}}\big)|v_{\text{g}} - v_{\text{p}}|\\
&\frac{dw_{\text{p}}}{dt}=\frac{3\rho_{\text{g}}C_{\text{D}}}{4D_{\text{p}}\rho_{\text{p}}}\big(w_{\text{g}} - w_{\text{p}}\big)|w_{\text{g}} - w_{\text{p}}| + g,
\end{align}
where $u_\text{p}$, $v_\text{p}$, and $w_\text{p}$ are the particle velocity in the $x$-, $y$-, and $z$-directions, respectively, $u_\text{g}$, $v_\text{g}$, and $w_\text{g}$ are the gas velocity in the $x$-, $y$-, and $z$-directions, respectively, $C_\text{D}$ is the drag coefficient, and $g$ is the acceleration due to gravity. The drag force is given by:
\begin{equation}
C_{\text{D}}=\begin{cases}
\frac{24}{\text{Re}} & \text{if Re}\le 1 \\
\frac{24}{\text{Re}}\big(1+0.15\text{Re}^{0.687}\big) & \text{if Re}>1,
\end{cases}
\end{equation}
where Re is the Reynolds number. The Reynolds number for this case is given by:
\begin{equation}
\text{Re}=\frac{|\vec{u}_{\text{p}}-\vec{u}_{\text{g}}|D_{\text{p}}}{\nu}, 
\end{equation}
where $\vec{u}_\text{p}=(u_\text{p},v_\text{p},w_\text{p})$ and $\vec{u}_\text{g}=(u_\text{g},v_\text{g},w_\text{g})$. In this study, we represent cases with co-flow only, such that $v_\text{g}=0$ and $w_\text{g}=0$. The gas velocity is predicted using the turbulent jet model of \citet{lee2003} by modeling the infectious individual's mouth as a circular orifice with diameter $D_{\text{mouth}}$. Under this model, the co-flow velocity in the flow establishment zone ($s\le6.2D_{\text{mouth}}$) is given by:
\begin{equation}
u_\text{g}=
\begin{cases}
u_0 & \text{if }r\le R_0\\\
u_0\exp{\bigg(-\frac{(r-R_0)^2}{b^2}\bigg)} & \text{if }r>R_0,
\end{cases}
\end{equation}
where $r$ is the radial distance from the jet center line\red{, b is the Gaussian half-width}, and $R_0=D_{\text{mouth}}/2 - s/12.4$\red{, where $s$ is the distance traveled along the jet center line}. In the zone of established flow ($s>6.2D_{\text{mouth}}$), the velocity is given by:
\begin{equation}
u_\text{g}=u_0\exp{\bigg(-\frac{r^2}{b^2}\bigg)}.
\end{equation}

The expelled air is typically warmer, moister, and contains more CO$_2$ than the background air \citep{mahyuddin2014}. The elevated temperature and water content cause the jet to curve upward due to buoyancy, which is partially offset by the increases in air density from the elevated concentration of CO$_2$. We follow the approach of \citet{baturin1972} to model the curve of the jet: 
\begin{equation}
z(x)=0.0354\sqrt{A_0}\text{Ar}_0\bigg(\frac{x}{\sqrt{A_0}}\bigg)^3\sqrt{\frac{T_0}{T_{\text{v},\infty}}},
\end{equation}
where $T_0$ is the source temperature, $T_{\text{v},\infty}$ is the background temperature, and $A_0$ is the cross-sectional area of the jet orifice. The \red{Archimedes number} Ar$_0$ is given by:
\begin{equation}
\text{Ar}_0=\frac{g\sqrt{A_0}}{u_0^2}\frac{\Delta\rho}{\rho_0},
\end{equation}
where $\Delta\rho=\rho_{\infty}-\rho_0$ is the difference between the background density, $\rho_{\infty}$, and the source density, $\rho_{0}$.

Similar to \citet{xie2007} and \citet{wei2015}, we use the relationships from \citet{chen1980} to represent the evolution in $T_{\text{v}}$ and the water vapor density $\rho_\text{v}$ within the turbulent jet. Once in the zone of established flow ($x\ge6.2D_{\text{mouth}}$), the center line temperature $T_{\text{v,c}}$ and the center line water vapor $\rho_{\text{v}}$ both decay with distance $s$ according to the same relationship:
\begin{equation}
    \frac{T_{\text{v,c}}-T_{\text{v},\infty}}{T_{\text{v},0}-T_{\text{v},\infty}} = \frac{\rho_{\text{v},c}-\rho_{\text{v},\infty}}{\rho_{\text{v},0}-\rho_{\text{v},\infty}} =
    \frac{5}{s/D_{\text{mouth}}}\sqrt{\frac{T_{\text{v},0}}{T_{\text{v},\infty}}}.
\end{equation}
In the flow establishment zone of the jet ($x<6.2D_{\text{mouth}}$), we assume that $T_{\text{v,c}}=T_{\text{v},0}$ and $\rho_{\text{v,c}}=\rho_{\text{v},0}$. 

The variation in $T_{\text{v}}$ and $\rho_{\text{v}}$ with the jet radius is then given by:
\begin{equation}
    \frac{T_{\text{v}}-T_{\text{v},\infty}}{T_{\text{v,c}}-T_{\text{v},\infty}} =
    \frac{\rho_{\text{v}}-\rho_{\text{v},\infty}}{\rho_{\text{v,c}}-\rho_{\text{v},\infty}} =
    \exp\bigg(-\frac{r^2\ln 2}{(0.11s)^2}\bigg).
\end{equation}

Water vapor is represented as an ideal gas, such that $\rho_{\text{v}}$ is computed from the partial pressure of water vapor $p_{\text{v}}$:
\begin{equation}
    \rho_{\text{v}}=\frac{M_{\text{w}}p_{\text{v}}}{RT_{\text{v}}}
\end{equation}

\begin{figure}
\centering
\includegraphics[width=6.2in]{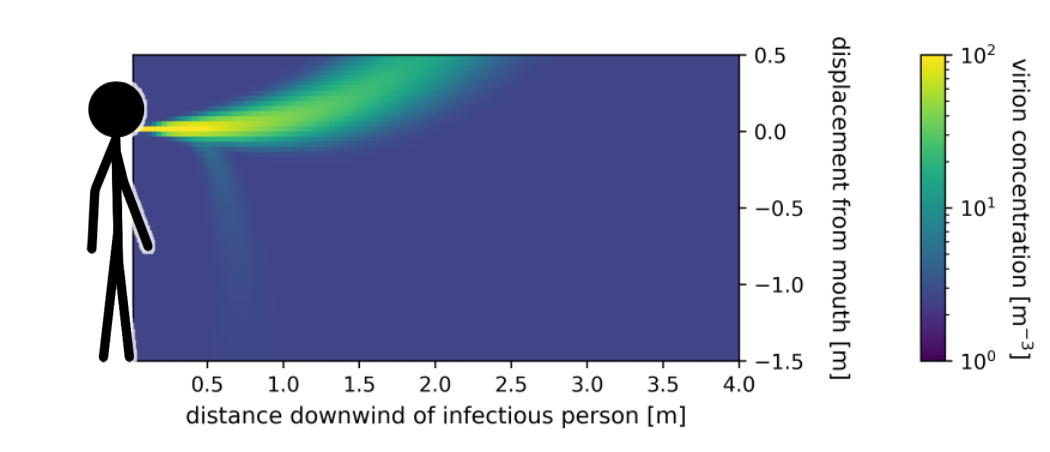}
\caption{Number concentration of SARS-CoV-2 virions resulting from continuous talking for one hour in an example case (see Table~\ref{tab:inputs} for input parameters). The virion concentration is one to two orders of magnitude greater in the expiratory jet than in the well-mixed room.}
\label{fig:scenario__C_vs_xz} 
\end{figure}

Using this combined jet dispersion model, we find that the spatial distribution in particles simulated by QuaRAD agrees well with the Discrete Random Walk (DRW) model of \citep{wei2015} (see comparison in Appendix~\ref{ap:verify}). Whereas the DRW model tracks thousands of Monte Carlo particles, QuaRAD simulates six quadrature points. The spatial distribution in virions associated with each QuaRAD point is shown in Fig.~\ref{fig:scenario__C_vs_xz}.

\begin{figure}
\centering
\includegraphics[width=3.in]{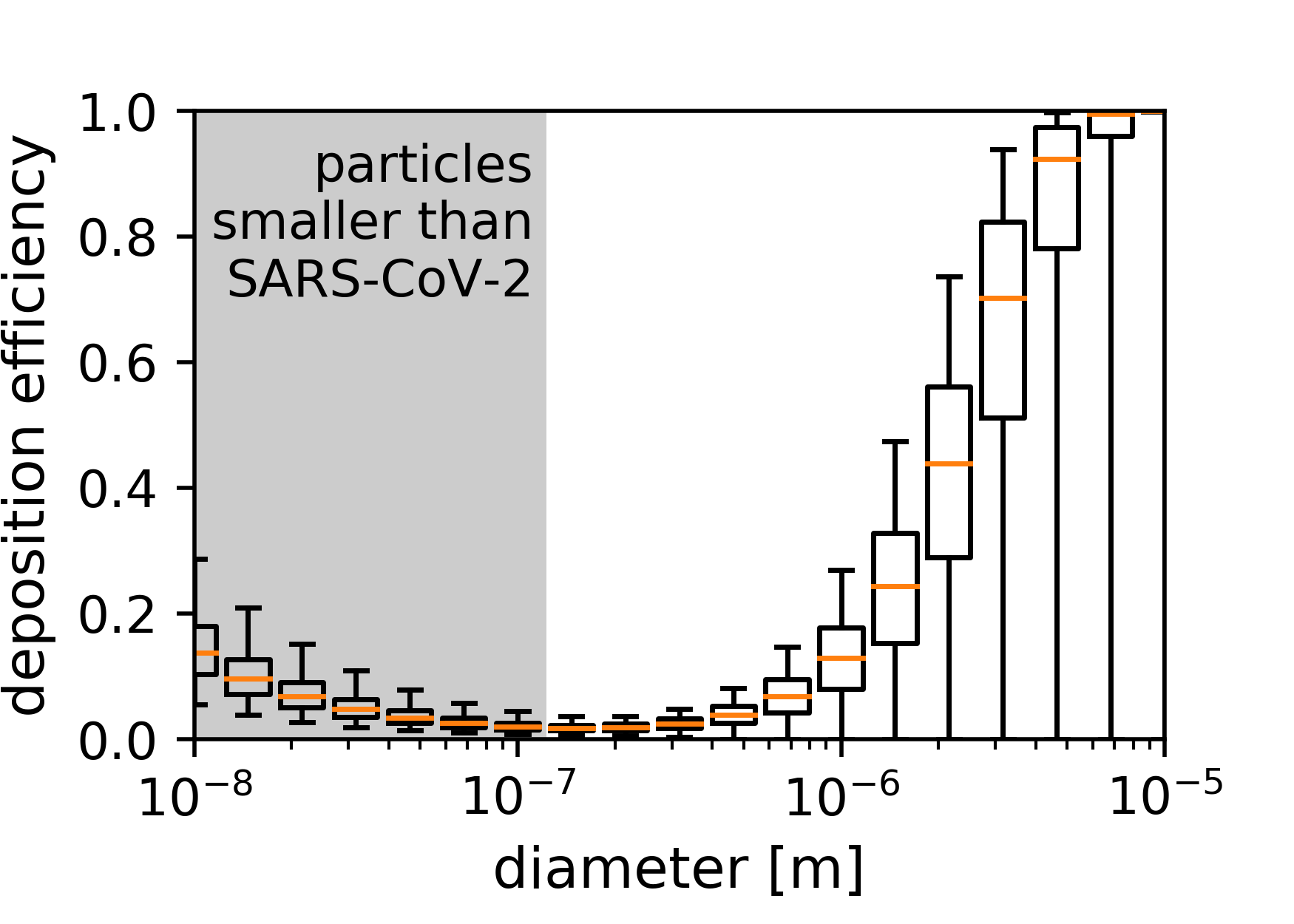}
\caption{The median (orange line), quartiles (boxes), and 95\% confidence intervals (whiskers) of the distribution in deposition efficiency as a function of particle size. The parameters within the deposition model (Eqn.~\ref{eqn:deposition}) were sampled according to the distributions in Table~\ref{tab:inputs}.}
\label{fig:deposition_efficiency} 
\end{figure}

\begin{figure}
\centering
\includegraphics[width=3.in]{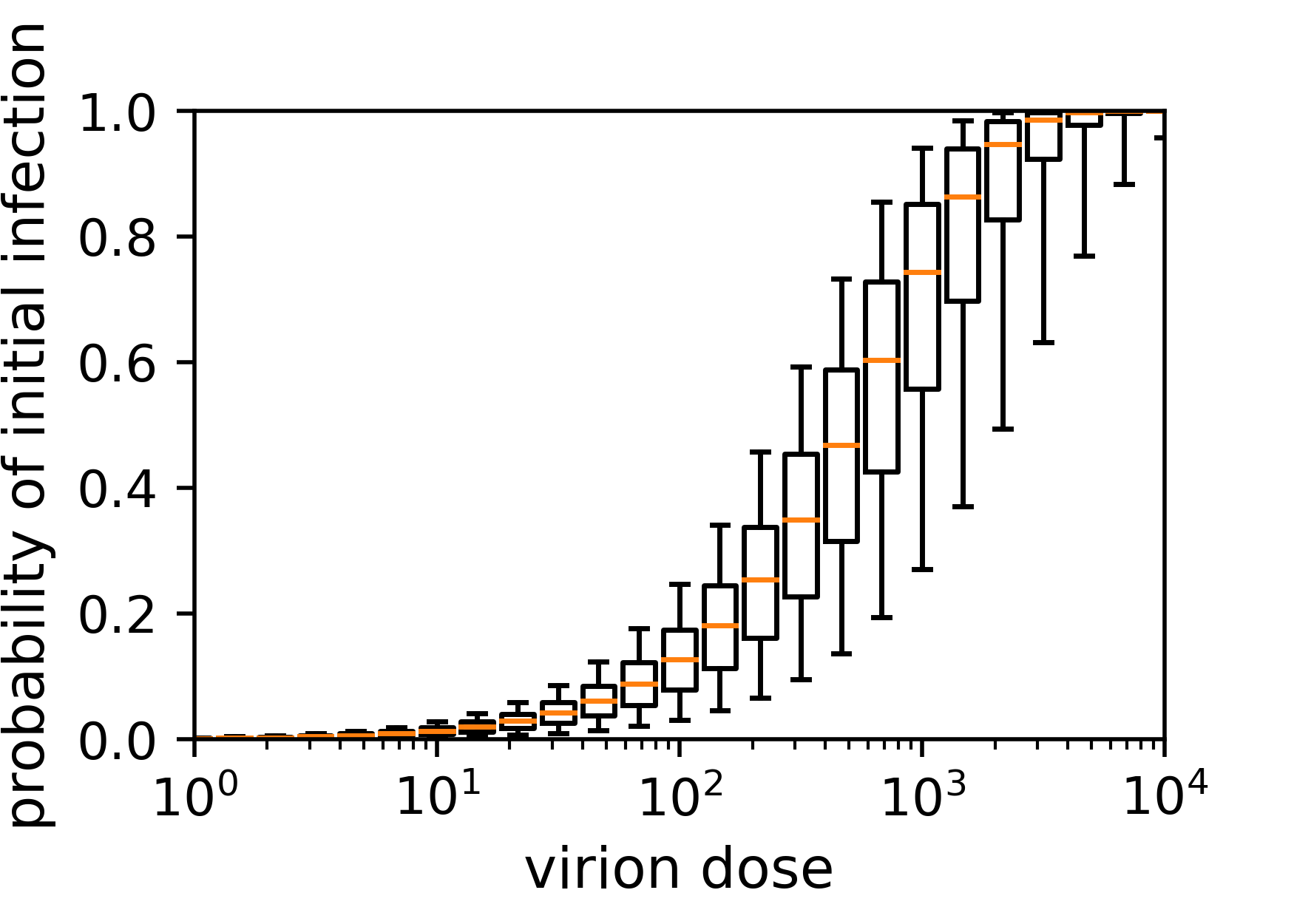}
\caption{The median (orange line), quartiles (boxes), and 95\% confidence intervals (whiskers) of the probability of initial infection given a SARS-CoV-2 virion dose. The parameters within the infection model (Eqns.~\ref{eqn:dose_response}) were sampled according to the distributions in Table~\ref{tab:inputs}.}
\label{fig:ensemble__dose_reseponse} 
\end{figure}

\subsection{Virion exposure and risk of infection}\label{sec:infection}
We applied the infection model described in \citet{gale2020} to quantify the probability of initial SARS-CoV-2 infection as a function of the virion exposure level and host defense through the mucus barrier, such that the probability of infection increase with virion exposure and decreases with virion binding affinity to mucin molecules. The greater the number of virions to which a person is exposed, $V_{\text{e}}$, the greater the number of chances that any given virion will successfully infect a cell. Differences in individuals' immune responses are represented by modifying the probability that a virion infects a single cell, $p_1$. The probability of initial infection within the host, $p_{\text{infect}}$, is then computed as a function of $p_1$ and $V_{\text{e}}$:
\begin{equation}\label{eqn:dose_response}
p_{\text{infect}}=1 - (1-p_1)^{V_{\text{e}}}.
\end{equation}

To compute $V_{\text{e}}$ from the virion concentrations simulated in QuaRAD, we applied a respiratory deposition model to calculate the number of inhaled virions that reach the potential site of infection. In the case of SARS-CoV-2, infection must begin with a virion binding to an angiotensin-converting enzyme 2, or ACE2, receptor \citep{lukassen2020, perrotta2020}. As ACE2 expression within the respiratory tract is highest in the nasal epithelium and this is the most likely site of initial infection \citep{hou2020, matheson2020, ziegler2020}, we assume that $V_{\text{e}}$ is the total number of virions that deposit to this region. We applied the model from \citet{cheng2003} to predict the deposition efficiency to the nasal epithelium as a function of particle size:
\begin{equation}\label{eqn:deposition}
e_{\text{d}} = 1 - \exp\bigg(aD_{\text{a}}^2(6\times10^4\dot{V}_{\text{breathe}}) + bD_{\text{diff}}^c(6\times10^4\dot{V}_{\text{breathe}})^d\bigg),
\end{equation} 
where $D_{\text{a}}$ is the aerodynamic diameter of the particle, $D_{\text{diff}}$ is the molecular diffusion coefficient of the particle, $\dot{V}_{\text{breathe}}$ is the breathing rate, and $a, b, c,$ and $d$ are coefficients of deposition efficiency. We represent the uncertainty reported by \citet{cheng2003} in the coefficients $a$, $b$, $c$, and $d$, leading to uncertainty in $e_{\text{d}}$ even for particles of the same size,. Fig.~\ref{fig:deposition_efficiency} shows the resulting variability in the size-dependent deposition represented across the ensemble of scenarios.

The overall rate at which virions deposit to the nasal epithelium, $dV_{\text{e}}/dt$, is computed as the sum over the deposition rates associated with each of the quadrature points:
\begin{equation}
\frac{dV_{\text{e}}}{dt}(x,y,z) = \dot{V}_{\text{breathe}}\sum_i^{N_{\text{quad}}}e_{\text{d},i}N_{\text{v},i}(x,y,z),
\end{equation}.

The probability that a virion infects a single cell, $p_1$, is  given by: 
\begin{equation}
    p_1=F_{\text{trans }}F_{\text{v }}F_{\text{c }}p_{\text{pfu }}p_{\text{cell}},
\end{equation}

where $F_{\text{trans}}$ is the fraction of virions \blue{that deposit to} the infection site, $F_{\text{v}}$ is the fraction of virions at the infection site but not bound to the mucin, $F_{\text{c}}$ is the fraction of virions already bound to other cells, $p_{\text{pfu}}$ is the probability a given virion is capable of initiating an infection, and $p_{\text{cell}}$ is the probability that a cell with a bound virus will become infected, defined as the virus successfully entering the cell, replicating, and releasing progeny virions. We assume that $p_{\text{cell}}=0.5$ and $F_{\text{c}}=1$ based on \citet{hoffmann2020} and \citet{hui2020}. \red{We also assume that the only possible infection site is the nasal epithelium, and thus set $F_{\text{trans}}=1$ for all virions depositing in this region}. The fraction of `free' virions not bound to the mucin, $F_{\text{v}}$, is modeled as:
\begin{equation}
F_{\text{v}}=\frac{1}{K_{\text{mucin}}}[\text{Muc}_{\text{free}}],
\end{equation}

where $K_{\text{mucin}}$ is the association constant for binding of the virions to the mucin molecules and [$\text{Muc}_{\text{free}}$] is the concentration of free mucin molecules remaining. These parameters likely vary widely among individuals, and the distributions in their values remains poorly constrained. Following the reasoning of \citet{gale2020} and the measurements of \citet{hou2020} and \citet{kesimer2017}, we assume that $K_{\text{mucin}}$ varies between 10$^3$ and 10$^6$ M$^{-1}$ and that $[\text{Muc}_{\text{free}}]=7.11\times10^{10} \pm 1.23\times10^{10}$.

\section{Results}\label{sec:results}
We used QuaRAD to quantify the risk of airborne transmission of SARS-CoV-2 during a face-to-face encounter between an infectious and a susceptible person. Since a significant portion of SARS-CoV-2 transmission has been attributed to presymptomatic and asymptomatic carriers \citep{johansson2021, he2020, byambasuren2020}, who may transmit the virus before symptoms develop or may never develop symptoms, we focused on particles that are expelled when an infectious person is speaking. We quantified the risk of \blue{initial} infection during a one-hour conversation as a function of distance between the infectious and the susceptible person, assuming that the infectious person is speaking continuously. To quantify uncertainty in transmission risk stemming from uncertainty in input parameters, we analyzed an ensemble of 10,000 simulations. First, we focus on an example scenario to illustrate how an infectious person influences virion concentrations within an indoor space.

\subsection{Spatial variation in virion concentrations}
A person shedding SARS-CoV-2 virions affects the average virion concentration in a room, but they have a far greater impact on virion concentrations in the region directly in front of them, as shown in Fig.~\ref{fig:scenario__C_vs_xz}. The overall concentration of virions, $N_{\text{v}}$, is the sum over the concentration of virions associated with each quadrature point, $N_{\text{v},i}$ for $i=1,..,N_{\text{quad}}$ (see Fig.~\ref{fig:scenario__Cq_vs_xz}). Since this scenario represents a well-ventilated space (5 air changes per hour), the virion concentration far from the infectious person is relatively low (1~virion per m$^{3}$). However, virion concentrations are orders of magnitude greater directly downwind of the infectious individual than in the rest of the room. A susceptible person standing in the expiratory jet of an infectious person will be exposed to a large concentration of virions, even \blue{if} they are in a well-ventilated---or, in principle, unenclosed---space.

\begin{figure}
\centering
\includegraphics[width=3.in]{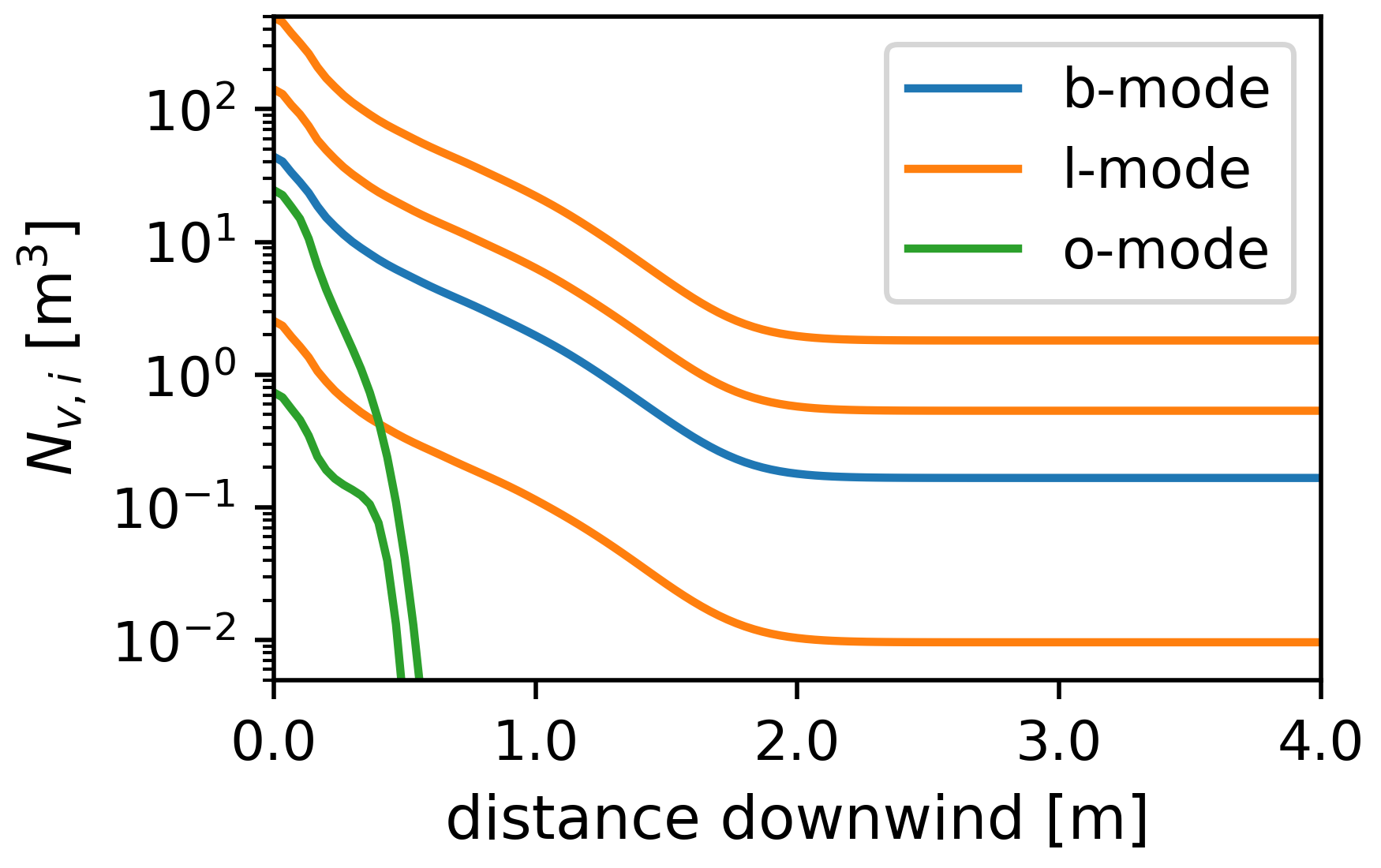}
\caption{Number concentration of virions $N_{\text{v},i}$ associated with each quadrature point $i=1,...,N_{\text{quad}}$. The values shown here correspond to slice at $z=0$ in Fig.~\ref{fig:scenario__Cq_vs_xz}, and the values at $x=0$ scale with the weights in Fig.~\ref{fig:quad_emission}c.}\label{fig:mode_contribution} 
\end{figure}

To identify the particle size ranges that are most important for virion transmission near and far from an infectious individual, we quantified the contribution of each quadrature point \blue{to} the overall virion concentration, shown as a function of downwind distance ($y=y_0$, $z=z_0$) in Fig.~\ref{fig:mode_contribution}. Virions tend to be concentrated in the fine (b- and l-mode) particles rather than the coarse (o-mode) particles, regardless of the distance between the infectious and susceptible individual. In this scenario, the contribution of virions in o-mode particles drops to negligible values beyond 0.5~m from the infectious person, reflecting the general pattern that o-mode particles quickly fall to the ground. On the other hand, the long residence times of b- and l-mode particles allow them to affect virion concentrations elsewhere in the well-mixed room, albeit in lower concentrations than in the expiratory jet of an infectious person, and to persist much longer within the jet itself. This near-field enhancement in airborne transmission from these fine particles in the b- and l-modes extends much farther than the reach of the larger o-mode particles.

\subsection{Near-field enhancement in the risk of airborne transmission}
For the full ensemble of 10,000 simulations, the absolute risk of initial infection is shown as a function of distance downwind of an infectious person ($y=y_0$, $z=z_0$) in Fig.~\ref{fig:short_range}a. The risk of an initial infection after an encounter with an infectious individual is orders of magnitudes higher when near the infectious individual than the average risk within the room. At any location, the risk of infection varies by over two orders of magnitude, due to the inherent variability in the parameters governing transmission. We caution that while this model is useful for comparing risk, predictions of absolute risk of initial infection (Fig.~\ref{fig:short_range}a) depend strongly on the assumed distribution in input parameters, many of which are poorly constrained.

At distances greater than three meters, the risk of initial infection by the susceptible person is equal to risk in a well-mixed room (assuming uniform concentration), whereas the risk of infection is orders of magnitude greater if the infectious and susceptible person are in a close, face-to-face conversation (0.5~m distancing). This enhancement in near-field risk is shown in Fig.~\ref{fig:short_range}b. Once the susceptible person moves beyond three meters from the infectious person, they avoid the near-field increases in virion concentrations and transmission is, thus, governed by the far-field concentration. These far-field concentrations are approximately the same as the concentration within a well-mixed room, reflected by values near unity for distances beyond three meters downwind in Fig.~\ref{fig:short_range}b. 

\subsection{Minimum distancing to avoid near-field enhancement in risk}
To quantify the distance that one needs to remain from a potentially infectious person to avoid near-field enhancements in risk, we identified the distance at which the near-field impact drops below 5\% of total transmission risk. Within this distance, local enhancements in virion concentrations increase transmission risk by more than 5\%. We find that this threshold distance, $x_{\text{near}}$, varies between one meter to four meters, suggesting large variability in near-field effects. In more than 50\% of cases, we find $x_{\text{near}}$ is greater than the common recommendation of two meters (six feet).

\begin{figure}
\centering
\includegraphics[width=3.5in]{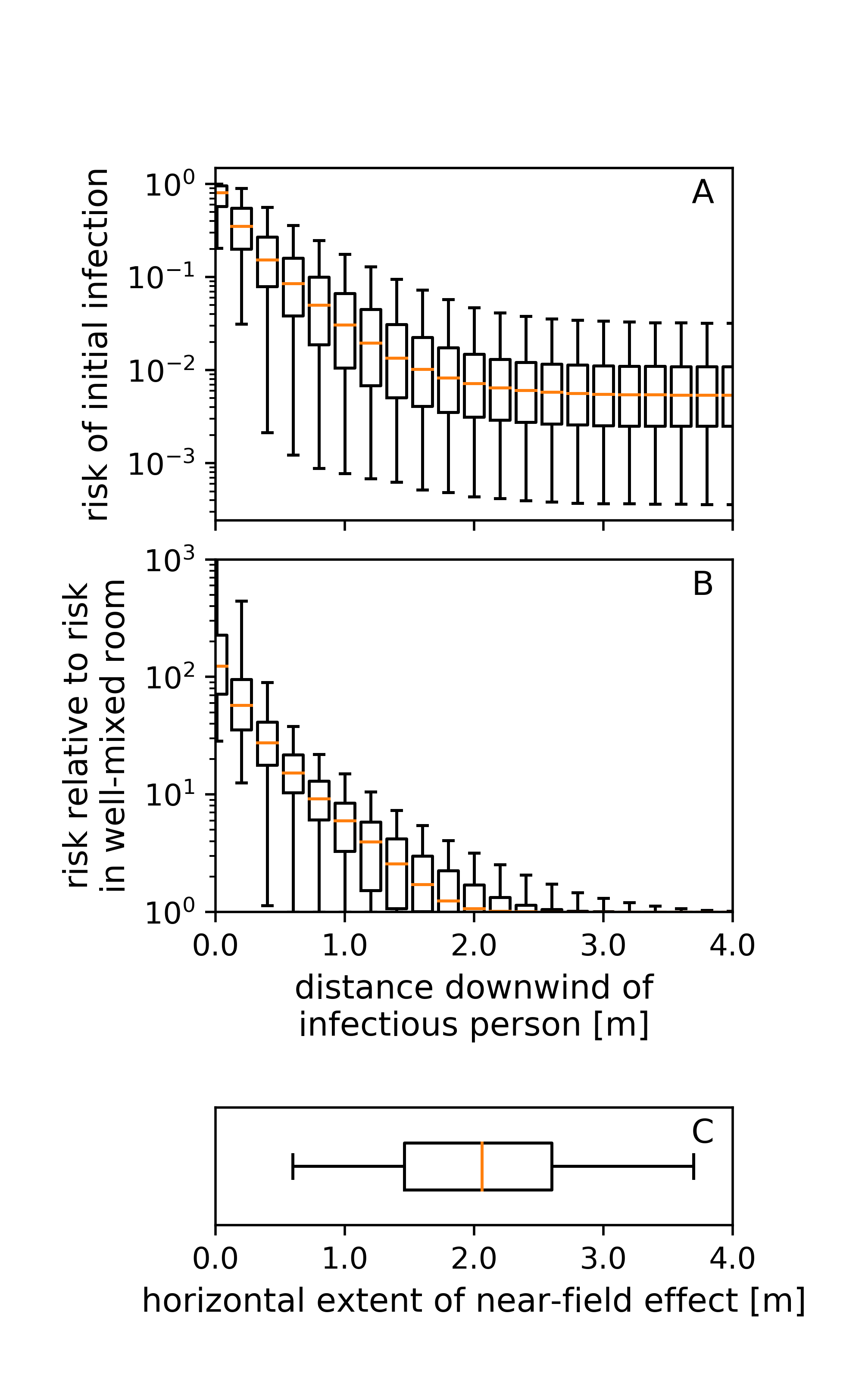}
\caption{The median (orange line), quartiles (boxes), and 95\% confidence intervals (whiskers) for (A) the absolute risk of initial infection and (B) enhancement in transmission risk relative to the risk in a well-mixed room. Each quantity is shown as a function of distance at which a susceptible person is standing downwind from an infectious individual. Across cases, we find wide variability in (C) distance wherein near-field enhancements strongly impact transmission; we define this threshold distance as the distance at which local enhancements account for 5\% of total transmission. All cases represent a person speaking at a medium volume for one hours.}
\label{fig:short_range} 
\end{figure}

To identify the model parameters that drive variability in $x_{\text{near}}$, we applied the Delta Moment-Independent Measure \citep{borgonovo2007} within the Sensitivity Analysis in Python (\url{https://salib.readthedocs.io/en/latest/}). The sensitivity index for each input parameter quantifies the fractional reduction in the variance of $x_{\text{near}}$ if the uncertainty in that parameter is eliminated. Across simulations, variability in $x_{\text{near}}$ is driven predominantly by variability in $u_0$, the velocity at which particles are expelled, with a sensitivity index of 0.9. On the other hand, the most poorly constrained parameters, such as the rate of viral shedding and parameters governing the immune response in the new host, did not strongly affect variability in $x_{\text{near}}$. Whereas the distribution in $u_0$ applied in this study represents an infectious person who is speaking at normal volume, the value of $u_0$ will be greater for a person who is sneezing, coughing, or speaking at a louder volume. For these events, we would then expect enhanced near-field transmission beyond the range of $x_{\text{near}}$ shown in Fig.~\ref{fig:short_range}c.

\section{Conclusions}\label{sec:conclusions}
This paper describes the development the Quadrature-based model of Respiratory Aerosol and Droplets (QuaRAD) for simulating the evolution of respiratory particles and its application to simulate near-field transmission of airborne viruses. We showed that the risk of airborne transmission is elevated by orders of magnitude in the expiratory jet of the infectious person. The horizontal extent of this near-field increase in transmission risk was highly variable among simulations, and this variability was controlled by the velocity of the expired particles. 

Since the risk of airborne transmission is strongly enhanced near infectious individuals, maintaining distance from those outside one's household is a highly effective measure for reducing the spread of airborne pathogens. If a susceptible person moves far enough away, they can avoid the local increase in virion concentrations within the infectious person's expiratory jet; at these long-range distances, the risk of infection is approximately the same as the risk within a well-mixed room. However, the simulations presented in this paper revealed that the elevated risk due to local increases in the virion concentration often extend beyond the typical two-meter (six-foot) distancing guidelines. These near-field impacts will extend even farther if the infectious person speaks loudly, sings, or even breathes heavily while exercising, as these activities expel particles at a higher velocities than conversational speech. \blue{On the other hand, if the infectious person is wearing a mask, they will expel particles at a slower velocity than if they are unmasked \citep{maher2020}, reducing the horizontal extent of their near-field influence.}

In general, we found tremendous variability of predicted transmission risk during any given encounter, due to the inherent variability in physiological properties of the infectious and susceptible individuals and in room conditions. To avoid the near-field enhancements in airborne exposure, distancing of at least three meters should be maintained. For situations in which distancing is not possible, such as interactions between healthcare workers and patients, it is important to provide protective equipment against airborne transmission, such as N95 respirators.

\section{Data Availability}
The QuaRAD source code, input files, and processing script are available for download at: \url{https://github.com/lfierce2/QuaRAD/}. Simulation ensembles were created using latin hypercube sampling with pyDOE: \url{https://pythonhosted.org/pyDOE/}. The sensitivity analysis was performed using the Sensitivity Analysis Library in Python, which is available at: \url{https://salib.readthedocs.io/en/latest/}.

\section{Acknowledgements}
This research was supported by the DOE Office of Science through the National Virtual Biotechnology Laboratory, a consortium of DOE national laboratories focused on response to COVID-19, with funding provided by the Coronavirus CARES Act. This project was supported in part by the U.S. Department of Energy through the Office of Science, Office of Workforce Development for Teachers and Scientists (WDTS) under the Science Undergraduate Laboratory Internships Program (SULI) and the Environmental and Climate Sciences Department under the BNL Supplemental Undergraduate Research Program (SURP). The quadrature-based model was originally developed for simulation of atmospheric aerosol with support from the DOE Atmospheric System Research program at Brookhaven National Laboratory, a multiprogram national laboratory supported by DOE Contract DE-SC0012704.

\clearpage
\begin{longtable}[h]{llc}
    \multicolumn{3}{ c }{ \textbf{Definition of Variables}} \\
    \hline
    \textbf{variable} & \textbf{definition} & \textbf{units} \\ \hline
    \endfirsthead
    
    \multicolumn{3}{ c }{ \textbf{Continued definition of Variables}} \\
    \hline
    \textbf{variable} & \textbf{definition} & \textbf{units} \\ \hline
    \endhead
    
    $A$ & surface area available for deposition & m$^2$ \\
    $A_{0}$ & cross-sectional area of the jet orifice & m$^{2}$ \\
    ACH & air changes per hour & h$^{-1}$ \\
    Ar$_{0}$ & Archimedes number & - \\
    $a,b,c,d$ & nasal deposition efficiency coefficients & - \\
    
    $b$ & Gaussian half-width of jet & m \\
    
    $C_{\text{D}}$ & drag coefficient & - \\
    $c_{i}$ & concentration associated with quadrature point $i$ & particles/m$^{3}$ \\

    $C_{\text{p}}$ & specific heat of particle & J/(K kg) \\
    $C_{\text{T}}$ & correction factor for diffusion coefficient & - \\
    
    $D_0$ & initial particle diameter & m \\
    $D_{\text{p}}$ & particle diameter & m \\
    $D_{\text{a}}$ & particle aerodynamic diameter & m \\
    $D_{\text{d}}$ & particle dry diameter & m \\
    $D_{\text{diff}}$ & molecular diffusion coefficient of particle & m$^2/\text{s}$ \\
    $D_{\infty}$ & diffusivity of water in air & $2.42\times10^{-5}$ m$^{2}/$s \\
    $D_{\text{mouth}}$ & diameter of infectious individual's mouth & m \\
    
    $e_{\text{d}}$ & deposition efficiency to nasal epithelium & - \\
    
    $F_{\text{c}}$ & fraction of virions in exposure dose already & - \\
     & bound to cells & \\
    $F_{\text{trans}}$ & fraction of virions in exposure dose transported & - \\
     & to the infection site & \\
    $F_{\text{v}}$ & fraction of virions in exposure dose not & - \\
     & bound to mucin & \\
    $f_{\text{v}}$ & fraction of virions residing in coarse particles & - \\
     
    $g$ & acceleration due to gravity & $9.81$ m/s$^{2}$ \\
    
    $H$ & height of the room ($z$-direction) & m \\
    $h_{i}$ & quadrature abscissas & - \\
    $H_n$ & Hermite polynomial & - \\
    
    $I_{\text{rate}}$ & average adult inhalation rate & breath/min \\
    $I_{\text{vol}}$ & average adult inhalation volume & m$^3$/breath \\
    
    $k$ & fresh air ventilation rate & s$^{-1}$ \\
    $k_{\text{g}}$ & thermal conductivity of air & $0.2529$ W/(m K) \\
    $K_{\text{mucin}}$ & association constant for binding between & M$^{-1}$ \\
     & virion and mucin & \\
    $\kappa$ & effective hygroscopicity parameter of aerosol & - \\
     & contained in particle & \\
    
    $L$ & length of the room ($x$-direction) & m \\
    $L_{\text{v}}$ & latent heat of vaporization & $2.45 \times 10^{6}$J/kg \\
    
    $m_{\text{p}}$ & mass of aqueous particle p & kg \\
    $M_{\text{w}}$ & molecular weight of water & $0.018$ kg/mol \\
    
    $\mu_{\text{b}}, \mu_{\text{l}}, \mu_{\text{o}}$ & Geometric mean diameter of particles in the & m \\
     & b-, l-, and o-mode, respectively & m \\
    
    $[\text{Muc}_{\text{free}}]$ & concentration of free mucin molecules & molecules/mm$^{3}$ \\
    
    $N_{\text{b}}, N_{\text{l}}, N_{\text{o}}$ & particulate number emission rate for the b-, l-, & particles/s \\
     & and o-mode, respectively & \\
    $N_{\text{deposit}}$ & rate of virion deposition into the nasal epithelium & virions/s \\
    $N_{\text{p}}$ & overall number concentration of particles & particles/m$^{3}$ \\
    $N_{\text{v}}$ & overall number concentration of virions & virions/m$^{3}$ \\
    $N_{\text{v},i}$ & virions associated with quadrature point $i$ & virions/m$^{3}$ \\
    Nu & the Nusselt number & - \\
    $\nu$ & dynamic viscosity & Pa s \\
    $\dot{N}_{\text{v}}$ & number emission rate of virions & virions/s \\
    
    $p$ & ambient pressure & Pa \\
    $p_{1}$ & probability that a single virion initiates an infection & - \\
    $p_{\text{cell}}$ & probability, given a bound virion, that a & $0.5$ \\
     & cell becomes infected & \\
    $p_{\text{infect}}$ & probability of initial infection given an exposure dose & - \\
    $p_{\text{pfu}}$ & probability that a given virion is capable of & - \\
     & initiating infection in a cell & \\
    $p_{\text{v},0}$ & saturation vapor pressure of the air & Pa \\
    $p_{\text{v},\infty}$ & vapor pressure far from droplet surface & Pa \\
    $p_\text{{v,p}}$ & vapor pressure at droplet surface & Pa \\
    Pr & the Prandtl number & - \\
    
    $R_0$ & radius of the jet's potential core & m \\
    $R$ & universal gas constant & $8.314$ J/(mol K) \\
    $r$ & radial distance from the jet center line & m \\
    $R_{x}, R_{y}, R_{z}$ & reflection terms in the $x$-, $y$-, and $z$-directions, & - \\
     & respectively & \\
    Re & the Reynolds number & - \\
    RH & relative humidity & \% \\
    
    $\rho_{0}$ & density of expired air & kg/m$^{3}$ \\
    $\rho_{\infty}$ & density of background air & kg/m$^{3}$ \\
    $\rho_{\text{aero}}$ & density of aerosol in particle & kg/m$^{3}$ \\
    $\rho_{\text{g}}$ & density of gas & kg/m$^{3}$ \\
    $\rho_{\text{p}}$ & density of particle & kg/m$^{3}$ \\
    $\rho_{\text{v,c}}$ & center line water vapor density & kg/m$^{3}$ \\
    $\rho_{\text{w}}$ & density of water & $1000$ kg/m$^{3}$ \\
    
    $S_{0}$ & initial plume saturation ratio & - \\
    $S_{\infty}$ & background saturation ratio & - \\
    $s$ & distance traveled along center line of the jet & m \\
    Sc & the Schmidt number & - \\
    Sh & the Sherwood number & - \\
    
    $\sigma_{\text{s/a}}$ & surface tension on particle surface & N/m \\
    $\sigma_{\text{b}}, \sigma_{\text{l}}, \sigma_{\text{o}}$ & geometric standard deviation of particle diameter & - \\
     & in the b-, l-, and o-mode, respectively & - \\
    
    $t$ & time & s \\
    $T_{0}$ & initial plume temperature & K \\
    $T_{\text{v},\infty}$ & background temperature & K \\
    $T_{\text{p}}$ & particle temperature & K \\
    $T_{\text{v}}$ & vapor temperature & K \\
    $T_{\text{v,c}}$ & centerline temperature & K \\
    
    $u_{0}$ & initial expiration velocity & m/s \\
    $u_{\text{g}}$ & gas velocity in the $x$-direction & m/s \\
    $u_{\text{p}}$ & particle velocity in the $x$-direction & m/s \\
    
    $V$ & room volume & m$^3$ \\
    $\dot{V}_{\text{breath}}$ & volumetric breathing rate & m$^3$ / s \\
    $V_{\text{e}}$ & number of virions in airborne exposure dose & virions \\
    $\text{Vf}_{\text{aero}}$ & bulk volume fraction of aerosol in droplet & m$^{3}/$m$^{3}$ \\
    $v_{\text{g}}$ & gas velocity in the $y$-direction & $0$ m/s \\
    $v_i$ & viral load associated with quadrature point $i$ & virions/m$^{3}$ \\
    $v_{\text{p}}$ & particle velocity in the $y$-direction & m/s \\
    
    $W$ & width of the room ($y$-direction) & m \\
    $w_{\text{d}}$ & deposition rate onto surfaces & m/s \\
    $w_{\text{g}}$ & gas velocity in the $z$-direction & $0$ m/s \\
    $w_{i}$ & weight associated with quadrature point $i$ & - \\
    $w_{\text{p}}$ & particle velocity in the $z$-direction & m/s \\
    
    $x_{0}$ & $x$ location of source & m \\
    $x_{\text{c}}$ & $x$ location of jet center line for each particle & m \\
    $x_{\text{near}}$ & threshold distance at which local enhancement in virion & m \\
     & concentration increases transmission risk by $> 5\%$ & \\
    $y_{0}$ & $y$ location of source & m \\    
    $y_{\text{c}}$ & $y$ location of jet center line for each particle & m \\
    $z_{0}$ & $z$ location of source (mouth height) & m \\
    $z_{\text{c}}$ & $z$ location of jet center line for each particle & m \\
    
    \hline
\caption{Each variable used in this paper, along with its definition, units, and (if treated as a constant) its value. \label{tab:def}}
\end{longtable}

\clearpage

{\setstretch{1.0}
\begin{longtable}{l|ccc|c|c}
\centering
    \textbf{variable} & \textbf{dist.} & \parbox{3cm}{\centering\linespread{0.85}\bf min/mean/ \\geom. mean} &
    \parbox{3cm}{\centering\linespread{0.85}\bf max/st. dev./ \\geom. st. dev} &  \textbf{nominal value} & \textbf{ref.} \\ \hline \hline \endfirsthead

    \textbf{variable} & \textbf{dist.} & \parbox{3cm}{\centering\linespread{0.85}\bf min/mean/ \\geom. mean} &
    \parbox{3cm}{\centering\linespread{0.85}\bf max/st. dev./ \\geom. st. dev} &  \textbf{nominal value} & \textbf{ref.} \\ \hline \hline \endhead

    \multicolumn{6}{c}{Expiration Parameters} \\ \hline
    $D_{\text{mouth}}$ & normal & $0.02$ & $2\times10^{-4}$ & $0.02$ & \\
    $S_{0}$ & normal & $1$ & $5\times10^{-3}$ & $1$ & 1 \\
    $T_{0}$ & normal & $310.15$ & $0.1$ & $310.15$ & \\ \hdashline
    $u_{0}$ & normal & $4$ & $2$ & $4$ & 2 \\
    $z_{0}$ & normal & $1.5$ & $0.07$ & $1.5$ & 3 \\ \hline
    
    \multicolumn{6}{c}{Size Distribution Parameters} \\ \hline
    $\mu_{\text{b}}$ & normal & $1.60 \times 10^{-6}$ & $2.56 \times 10^{-7}$ & $1.60 \times 10^{-6}$ & \\ 
    $\mu_{\text{l}}$ & normal & $2.50 \times 10^{-6}$ & $3.75 \times 10^{-7}$ & $2.50 \times 10^{-6}$ & 4 \\ 
    $\mu_{\text{o}}$ & normal & $1.45 \times 10^{-4}$ & $1.16 \times 10^{-6}$ & $1.45 \times 10^{-4}$ & \\ \hdashline
    
    $N_{\text{b}}$ & normal & $6.75$ & $1.08$ & $6.75$ & \\
    $N_{\text{l}}$ & normal & $8.55$ & $1.28$ & $8.55$ & 4 \\
    $N_{\text{o}}$ & normal & $1.58 \times 10^{-3}$ & $1.3 \times 10^{-3}$ & $1.58 \times 10^{-3}$ & \\ \hdashline
    
    $\sigma_{\text{b}}$ & normal & $1.30$ & $0.02$ & $1.30$ & \\
    $\sigma_{\text{l}}$ & normal & $1.66$ & $0.05$ & $1.66$ & 4 \\
    $\sigma_{\text{o}}$ & normal & $1.80$ & $0.01$ & $1.80$ & \\ \hline
    
    \multicolumn{6}{c}{Aerosol Parameters} \\ \hline
    $\kappa$ & uniform & $0.3$ & $1.2$ & $1.1$ & 5 \\
    $\rho_{\text{aero}}$ & uniform & $1000$ & $1600$ & $1300$ & 6 \\
    $\text{Vf}_{\text{aero}}$ & uniform & $0.01$ & $0.09$ & $0.05$ & 7 \\ \hdashline
    
    $\dot{N}_{\text{v}}$ & lognormal & $7$ & $1.2$ & $7$ & 8 \\
    $f_{\text{v}}$ & lognormal & $3.60 \times 10^{-2}$ & $1.2$ & $3.60 \times 10^{-2}$ & \\ \hline
    
    \multicolumn{6}{c}{Room Parameters} \\ \hline
    \red{ACH} & uniform & $0.3$ & $2.7$ & $1.5$ & 9 \\
    $H$ & normal & $2.74$ & $0.38$ & $2.74$ & 10 \\
    $S_{\infty}$ & uniform & $0.25$ & $0.6$ & $0.5$ & 11 \\
    $T_{\text{v},\infty}$ & normal & $293.4$ & $2$ & $293.15$ & 12 \\ \hdashline
    $L$ & uniform & $7$ & $15$ & $10$ & 13 \\
    $W$ & uniform & $7$ & $15$ & $10$ & \\
    \hline
    
    \multicolumn{6}{c}{Inhalation \& Deposition Parameters} \\ \hline
    $I_{\text{rate}}$ & uniform & $12$ & $20$ & $16$ & 14 \\
    $I_{\text{vol}}$ & uniform & $3.75\times10^{-4}$ & $6.25\times10^{-4}$ & $4.69\times10^{-4}$ & 15 \\ \hdashline
    
    $a$ & normal & $-3.9\times10^{-3}$ & $2.33\times10^{-3}$ & $-3.9\times10^{-3}$ &  \\
    $b$ & normal & $-16.6$ & $4.5$ & $-16.6$ & 16 \\
    $c$ & normal & $0.5$ & $0.02$ & $0.5$ &  \\
    $d$ & normal & $-0.28$ & $0.09$ & $-0.28$ &  \\ \hline
    
    \multicolumn{6}{c}{Infection Parameters} \\ \hline
    $[\text{Muc}_{\text{free}}]$ & normal & $1.18\times10^{-7}$ & $2.04\times10^{-8}$ & $1.18\times10^{-7}$ & 17 \\ \hdashline
    $K_{\text{mucin}}$ & lognormal & $1000$ & $2$ & $1000$ & \\
    $p_{\text{cell}}$ & uniform & $0.1$ & $0.9$ & $0.1$ & 18 \\
    $p_{\text{pfu}}$ & lognormal & $2.8\times10^{-3}$ & $1.2$ & $2.8\times10^{-3}$ & \\ \hline
    
\caption{Sample distributions for each varied parameter, using minimum and maximum for uniform distributions, mean and standard deviation for normal distributions, and geometric mean and geometric standard deviation for lognormal distributions. Nominal values are those used when that parameter is not varied. \\
\footnotesize{References: 1. \citet{wei2015}, 2. \citet{chao2009,tang2013}, 3. \citet{nchs2021}, 4. \citet{johnson2011}, 5. \citet{petters2007,vejerano2018}, 6. \citet{stadnytskyi2020}, 7. \citet{vejerano2018}, 8. \citet{milton2013}, 9. \citet{epa2018,bennett2012,turk1987}, 10. \citet{gsa2019}, 11. \citet{nguyen2014}, 12. \citet{antretter2010}, 13. \citet{nystate2010}, 14. \citet{flenady2017}, 15. \citet{flenady2017, sidebotham2007}, 16. \citet{cheng2003}, 17. \citet{gale2020,kesimer2017}, 18. \citet{gale2020}.}\label{tab:inputs}}    
\end{longtable}}

\appendix
\counterwithin{figure}{section}
\section{Appendix}
\subsection{Quadrature optimization}\label{ap:quad_optimization}
\begin{figure}
    \centering
    \includegraphics[width=3.5in]{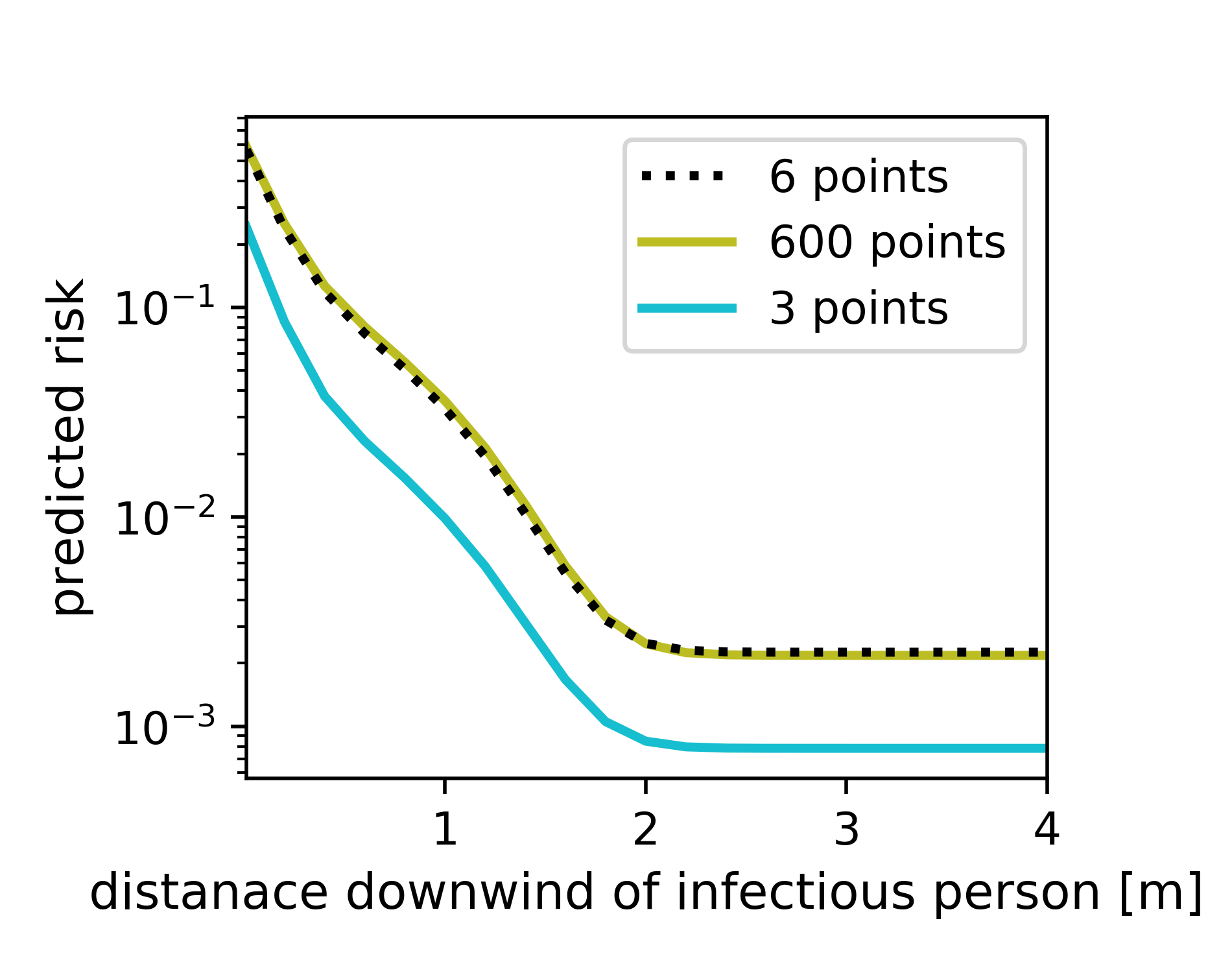}
    \caption{Comparison between optimized quadrature approximation applied in QuaRAD (6 points, dashed line) and the same simulation if more quadrature points are used than needed (600 points, green line) or if not enough quadrature points are used (3 points, teal line).}
    \label{fig:quad_optimization}
\end{figure}

The size distribution of respiratory particles is represented in QuaRAD using a total of six quadrature points --- 1-point, 3-point, and 2-point Gauss-Hermite quadrature for the b-, l-, and o-modes, respectively. We find that further increases in the number of quadrature points did not improve predictions; this is shown shown through comparison between the 6-point QuaRAD representation (blacked dashed line in Fig.~\ref{fig:quad_optimization}) and the same simulations but using 600 quadrature points (green line). On the other hand, if we use only one quadrature point for each of the b-, l-, and o-modes (teal line), we find the model accuracy decreases. 

\subsection{Verification against established Monte Carlo model}\label{ap:verify}
\begin{figure}
    \centering
    \includegraphics[width=3.2in]{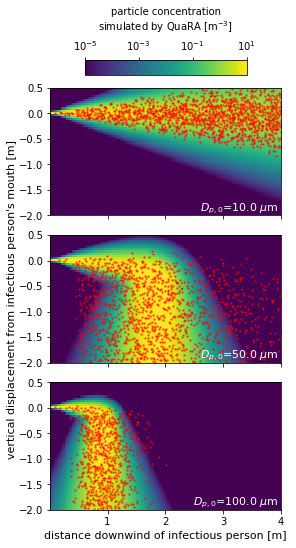}
    \caption{Comparison between particle concentrations predicted by QuaRAD using three quadrature points (color plot) and instantaneous particle distributions predicted by a DRW model tracking $50,000$ particles (scatter plot) \cite{wei2015} under the following conditions: $D_{\text{p}}=10$~$\mu$m, $D_{\text{p}}=50$~$\mu$m, and $D_{\text{p}}=100$~$\mu$m; 33 particles expelled per second in each case, $u_{0} = 10$ m/s, $T_{\text{v},\infty} = 298.15$ K, $T_{0} = 308.25$ K, and $S_{0} = 1$.}
    \label{fig:wei_verify}
\end{figure}

We verified particle dispersion simulated by QuaRAD through comparison with the Discrete Random Walk (DRW) model from \citet{wei2015}. For each particle size, the DRW tracks the evolution of thousands of Monte Carlo particle. Instantaneous snap shots of particles simulated by DRW are compared with the steady-state solution from the jet-puff model (Section~\ref{sec:dispersion}) in Fig.~\ref{fig:wei_verify}. The top, middle, and center panels in Fig.~\ref{fig:wei_verify} show predictions for particles of diameter $D_{\text{p}}=10$~$\mu$m, $D_{\text{p}}=50$~$\mu$m, and $D_{\text{p}}=100$~$\mu$m, respectively. In order to compare with the DRW simulations, evaporation and buoyancy were neglected in QuaRAD, and the particle expiration rate was adjusted to 33 particles per second. The steady-state concentration profiles simulated with QuaRAD agree with the DRW model (Fig.~\ref{fig:wei_verify}), particularly in the cases of 10~$\mu$m and 100~$\mu$m particles; given the predominance of particles smaller than 10~$\mu$m in the b- and l-modes and larger than 100~$\mu$m in the o-mode, these sizes are the most relevant to the dispersion dynamics in QuaRAD.

\clearpage

\bibliography{refs}

\end{document}